\documentclass[conference]{IEEEtran}
\IEEEoverridecommandlockouts

\usepackage{graphicx}
\usepackage{amsmath,amsfonts}
\usepackage{cite}
\usepackage{hyperref}

\usepackage{tikz}
\usetikzlibrary{positioning,fit,backgrounds}
\title{
Bayesian Changepoint Detection for Smart Sensing of Battery Degradation:\\
Cycle-Level Health Indicators and PyMC Implementation\thanks{Partially funded by the European Union. Views and opinion expressed are however those of the author(s) only and do not necessarily reflect those of the European Union or Europe’s Rail Joint Undertaking. Neither the European Union nor the granting authority can be held responsible for them. The project FP3-IAM4Rail is supported by the Europe's Rail Joint Undertaking and its members. Partially funded by AGH's subvention for scientific research.}
}

\author{
\IEEEauthorblockN{Waldemar Bauer}
\IEEEauthorblockA{Department of Automatic Control \& Robotics\\AGH University of Krakow\\Krakow, Poland\\
Email: bauer@agh.edu.pl\\
ORCID: 0000-0002-8543-0995}
\and
\IEEEauthorblockN{Anna Jarosz-Kozyro}
\IEEEauthorblockA{Department of Process Control\\
AGH University of Krakow\\
Krakow, Poland\\
Email: anjarosz@agh.edu.pl\\
ORCID: 0000-0002-7832-8191}
\and
\IEEEauthorblockN{Jerzy Baranowski}
\IEEEauthorblockA{Department of Automatic Control \& Robotics\\AGH University of Krakow\\Krakow, Poland\\
Email: jb@agh.edu.pl\\
ORCID: 0000-0003-3313-581X}
}

\begin{document}
\maketitle

\begin{abstract}
Reliable detection of the onset of accelerated degradation is central to safe and cost-efficient operation of lithium-ion batteries. This paper presents a Bayesian single-changepoint model applied to a simple but physically meaningful cycle-level health indicator (HI), defined as the ratio of charge time to discharge time. The indicator is computed directly from voltage–current telemetry typically available in battery management systems (BMS), without access to raw waveforms. The changepoint model is implemented in PyMC using Hamiltonian Monte Carlo and produces posterior distributions for onset time and pre/post-degradation slopes, together with posterior predictive checks. Experiments on an open 18650-cell remaining useful life (RUL) dataset show consistent midlife changepoints with narrow highest-density intervals. The formulation is lightweight, interpretable, and amenable to smart-sensing deployment on embedded BMS platforms.
\end{abstract}


\section{Introduction}

Lithium-ion batteries (LIBs) are widely used in electric vehicles, micro-mobility, and stationary storage. Their degradation affects internal resistance, usable capacity, power capability, and thermal behavior, all of which impact safety and lifecycle cost \cite{Berecibar2016,Visser2016}. Modern BMS architectures continuously sense voltage, current, and temperature (V–I–T) and often log cycle-level summary statistics. The challenge is to transform these routine measurements into robust, interpretable indicators of degradation and actionable maintenance decisions.

Many recent works use supervised machine learning or deep neural networks for state-of-health (SoH) and remaining useful life (RUL) prediction \cite{Li2020,GPR2018}. While such models can achieve high predictive accuracy, they often require large labeled datasets and high-rate signals, and they typically provide point estimates without calibrated uncertainty. For safety-critical systems, the absence of uncertainty quantification and the difficulty of interpreting black-box predictions limit their direct use in BMS decision logic.

Bayesian methods provide an attractive alternative: they offer principled uncertainty quantification, allow incorporation of prior information, and yield interpretable posterior summaries \cite{Wang2019}. In particular, \emph{changepoint} models explicitly describe transitions between regimes, such as the onset of accelerated degradation. Classical Bayesian changepoint formulations date back to \cite{Barry1993}, but applications to cycle-level LIB indicators that are compatible with embedded BMS constraints remain relatively limited.

This work proposes a simple, deployable framework in which: (i) a cycle-level health indicator (HI) is constructed from charge and discharge time; (ii) a Bayesian single-changepoint model is used to detect the onset of accelerated degradation in this HI; and (iii) implementation is carried out in PyMC using Hamiltonian Monte Carlo, with posterior predictive checks and diagnostics. A more extensive journal version is described in \cite{Jarosz2025Journal}; here we emphasize the Bayesian formulation, PyMC implementation, and smart-sensing implications.

Beyond purely empirical approaches, model-based statistical formulations have gained attention in the battery analytics community, especially those grounded in Bayesian inference. Recent surveys highlight how probabilistic frameworks naturally accommodate sensor uncertainty, device-to-device variability, and context-dependent degradation pathways \cite{Wang2019,Liu2020}. Bayesian models allow the incorporation of physical insight—for example, priors reflecting expected monotonic aging or plausible ranges of slopes—and facilitate downstream decision-making through posterior predictive distributions rather than point predictions. Similar ideas underlie Gaussian process–based SOH estimation \cite{GPR2018} and hierarchical models for population-level aging \cite{Berecibar2016}. However, few works explicitly focus on changepoint detection using low-dimensional cycle-level indicators, despite the practical importance of identifying the onset of accelerated wear rather than predicting SOH trajectories far into the future.

The changepoint detection perspective aligns directly with smart-sensing requirements: BMS engineers often care less about long-horizon RUL forecasting and more about reliably identifying transitions into regimes requiring derating, thermal safeguards, or warranty interventions. Our method therefore positions itself not as an alternative full-scale prognostic algorithm, but as a lightweight, interpretable sensing primitive that could be integrated into broader battery analytics pipelines. Furthermore, by using only V–I–T–derived cycle durations, the approach avoids the need for specialized hardware such as impedance spectroscopy units \cite{EIS2024Review} or fiber-optic sensing modules \cite{Fiber2024Review}, although such signals could be incorporated into extended models.

\section{Cycle-Level Health Indicator}

Production BMS often record, for each cycle $i$, the duration of charge and discharge phases. Let
\[
T^{\text{chg}}_i = \text{charging time (s)}, \qquad
T^{\text{dis}}_i = \text{discharging time (s)}.
\]
As a cell ages, internal resistance increases and usable capacity decreases. For a fixed protocol, this typically leads to longer charging times (e.g., slower constant-voltage taper) and shorter discharge times. We exploit this by defining the health indicator
\begin{equation}
\mathrm{HI}_i = \frac{T^{\text{chg}}_i}{T^{\text{dis}}_i}.
\label{eq:HI}
\end{equation}

This choice has several advantages:
\begin{itemize}
    \item It is physically interpretable: it compares energy input duration to energy output duration.
    \item It is directly computable from BMS logs without access to raw current or voltage waveforms.
    \item It is relatively robust to noise, as it aggregates over a full cycle rather than focusing on pointwise measurements.
\end{itemize}

\begin{figure}[t]
\centering
\includegraphics[width=\linewidth]{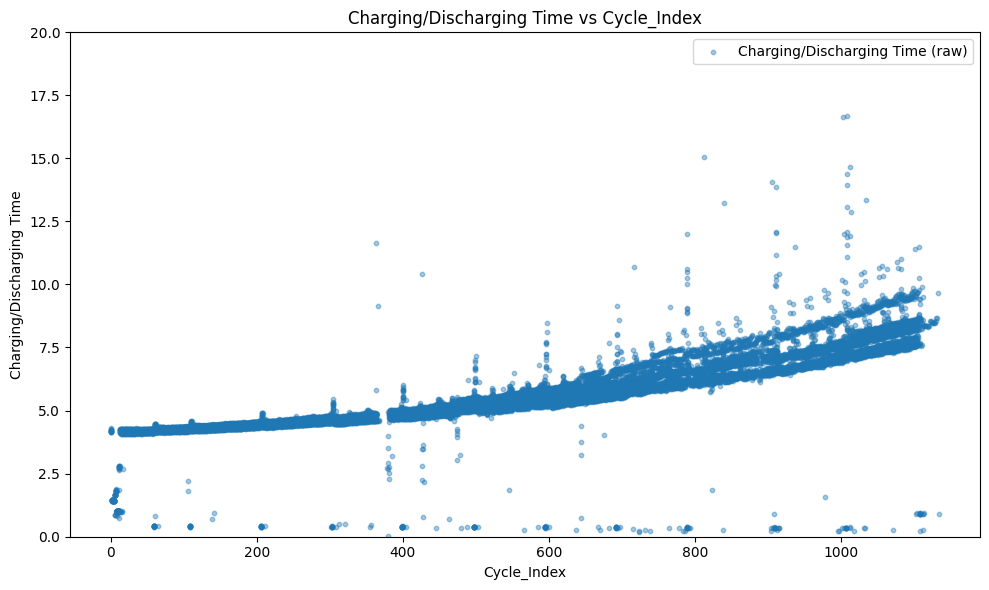}
\caption{Health indicator $\mathrm{HI}_i = T^{\text{chg}}_i / T^{\text{dis}}_i$ as a function of cycle index for all cells in the Battery RUL dataset \cite{BatteryRUL2022}. A midlife transition is visible in each trajectory.}
\label{fig:HI_timeseries}
\end{figure}

We evaluated this indicator using the “Battery Remaining Useful Life (RUL)” dataset \cite{BatteryRUL2022}, which provides cycle-level summaries for 14 commercial 18650 cells. Figure~\ref{fig:HI_timeseries} shows the HI across all cells. A characteristic pattern appears: an initial region with slowly varying HI, followed by a pronounced change in trend around midlife (roughly 350–450 cycles for most cells). This motivates a single-changepoint model.

In practice, alternative indicators (e.g., based on voltage plateaus or internal resistance estimates) could be used. However, (\ref{eq:HI}) has the advantage of requiring only two scalar durations per cycle, making it attractive for embedded systems.

\section{Bayesian Changepoint Model}

\subsection{Data Model}

Let $y_i = \mathrm{HI}_i$ denote the indicator at cycle $i$, and $t_i$ the corresponding cycle index. Because $y_i$ is positive and may exhibit multiplicative variation, we work with the log-transformed response:
\[
z_i = \log(y_i).
\]
We also standardize the cycle index to improve conditioning:
\[
\tilde{t}_i = \frac{t_i - \bar{t}}{s_t},
\]
where $\bar{t}$ and $s_t$ are the empirical mean and standard deviation of $t_i$.

We assume a piecewise-linear Gaussian model with a single unknown changepoint $\tau$ in the standardized domain:
\begin{align}
z_i &\sim \mathcal{N}(\mu_i, \sigma^2), \label{eq:lik}\\
\mu_i &=
\begin{cases}
\alpha_1 + \beta_1 \tilde{t}_i, & \tilde{t}_i < \tau, \\
\alpha_2 + \beta_2 \tilde{t}_i, & \tilde{t}_i \ge \tau.
\end{cases}
\label{eq:mu}
\end{align}
The parameter $\tau$ indicates the onset location; $(\alpha_1,\beta_1)$ describe the pre-change trend; $(\alpha_2,\beta_2)$ describe the post-change trend; and $\sigma$ is the residual standard deviation.

To clarify the probabilistic structure of the changepoint formulation, Fig.~\ref{fig:bn_changepoint} provides the corresponding Bayesian network, showing how global parameters, the changepoint~$\tau$, and standardized cycle indices jointly generate the per-cycle latent mean and the observed health-indicator values.

\subsection{Prior Specification}

We place weakly informative priors on all parameters. Let $\bar{z}$ and $s_z$ be the empirical mean and standard deviation of $z_i$. We use
\begin{align}
\tau &\sim \mathrm{Uniform}(\min \tilde{t}, \max \tilde{t}), \label{eq:prior_tau}\\
\alpha_1, \alpha_2 &\sim \mathcal{N}(\bar{z}, 5 s_z), \\
\beta_1, \beta_2 &\sim \mathcal{N}(0, 1), \\
\sigma &\sim \mathrm{HalfNormal}(s_z). \label{eq:prior_sigma}
\end{align}

These priors encode broad prior uncertainty: the intercepts are centered near the observed mean of $z_i$; slopes are initially assumed modest; and the changepoint is a priori equally likely across the observed range. The HalfNormal prior on $\sigma$ regularizes variance without imposing strong constraints.

The Bayesian formulation offers several advantages beyond merely detecting a structural transition. First, it enables coherent uncertainty quantification: instead of relying on thresholds applied to smoothed curves, which are sensitive to filter parameters, the posterior of the changepoint $\tau$ directly captures the probability distribution over plausible onset times. Second, Bayesian inference naturally handles uneven sampling, heteroscedastic noise, and missing cycles—conditions common in field BMS logs. Third, the piecewise-linear mean structure in (\ref{eq:mu}) is simple yet flexible enough to approximate many aging patterns observed in lithium-ion cells, including the emergence of the characteristic midlife "knee" \cite{Visser2016}.

Unlike classical changepoint estimators based on dynamic programming or penalized likelihoods, the Bayesian model treats $\tau$ as a continuous latent parameter rather than restricting it to discrete cycle indices. This continuous parametrization improves interpretability and improves mixing when modern gradient-based samplers such as NUTS \cite{hoffman2014nuts} are used. It also permits extension to hierarchical or multi-cell models where the changepoint is shared or partially pooled across units—an important feature for pack-level sensing.

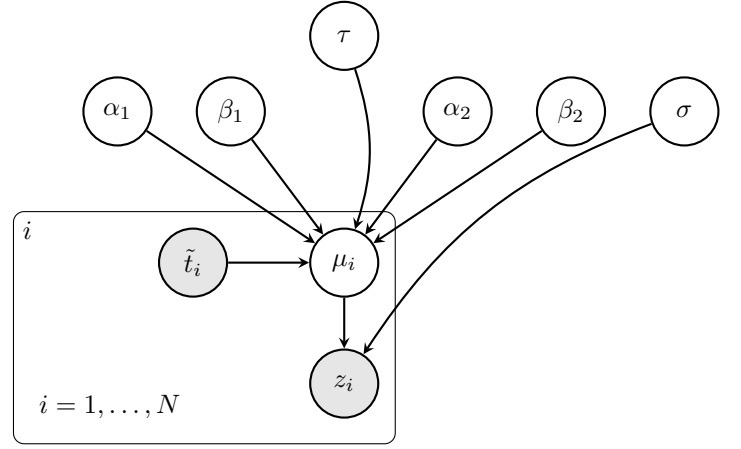
\begin{figure}[t]
\centering
\begin{tikzpicture}[
  latent/.style={circle,draw,minimum size=9mm},
  obs/.style={circle,draw,fill=gray!20,minimum size=9mm},
  const/.style={rectangle,draw=none},
  plate/.style={draw,rounded corners,inner sep=6pt},
  >=stealth,thick
]

\node[latent] (alpha1) at (-3,2) {$\alpha_1$};
\node[latent] (beta1)  at (-1.5,2) {$\beta_1$};
\node[latent] (tau)    at (0,3) {$\tau$};
\node[latent] (alpha2) at (1.5,2) {$\alpha_2$};
\node[latent] (beta2)  at (3,2) {$\beta_2$};
\node[latent] (sigma)  at (4.5,2) {$\sigma$};

\node[obs]    (ti)  at (-2,0) {$\tilde t_i$};
\node[latent] (mu)  at (0,0) {$\mu_i$};
\node[obs]    (z)   at (0,-1.6) {$z_i$};
\node[const]  (index) at (-3.1,-1.9) {$i=1,\dots,N$};

\draw[->] (alpha1) -- (mu);
\draw[->] (beta1)  -- (mu);
\draw[->] (alpha2) -- (mu);
\draw[->] (beta2)  -- (mu);
\draw[->] (tau) edge[bend left=18] (mu);
\draw[->] (ti) -- (mu);

\draw[->] (mu) -- (z);
\draw[->] (sigma) edge[bend right=18] (z);

\begin{scope}[on background layer]
  \node[plate,fit=(ti)(mu)(z)(index)] (platei) {};
\end{scope}
\node[const,anchor=north west] at (platei.north west) {$i$};

\end{tikzpicture}
\caption{Bayesian network for the single-changepoint model: global parameters
$(\alpha_1,\beta_1,\alpha_2,\beta_2,\tau,\sigma)$ govern the latent mean
$\mu_i$ and the observed health indicator $z_i$ over cycles with standardized
index $\tilde t_i$.}
\label{fig:bn_changepoint}
\end{figure}

The Bayesian network shown in Fig.~\ref{fig:bn_changepoint} highlights this structural simplicity: a handful of global parameters define the generative process for each cycle, ensuring that inference remains efficient even when long trajectories are available.

\subsection{Posterior and Inference}

Let $\theta = (\alpha_1,\beta_1,\alpha_2,\beta_2,\tau,\sigma)$ denote all parameters, and $z = (z_1,\dots,z_n)$ the observed data. The joint posterior is, up to a normalizing constant,
\begin{equation}
p(\theta \mid z) \propto
\left[\prod_{i=1}^n \mathcal{N}(z_i \mid \mu_i(\theta), \sigma^2)\right]
p(\theta),
\end{equation}
where $p(\theta)$ is the product of priors (\ref{eq:prior_tau})–(\ref{eq:prior_sigma}). Because the changepoint $\tau$ appears inside the piecewise definition (\ref{eq:mu}), the posterior is non-conjugate and multimodality is possible in general. We therefore resort to Markov chain Monte Carlo (MCMC) to approximate $p(\theta \mid z)$.


\section{PyMC Implementation}

\subsection{Modeling Workflow}

The model is implemented in PyMC (v5), a modern probabilistic programming library for Python that supports gradient-based MCMC via the No-U-Turn Sampler (NUTS) \cite{Salvatier2016}. The modeling workflow consists of:

\begin{enumerate}
    \item \textbf{Preprocessing:} compute $y_i$, $z_i$, and $\tilde{t}_i$; filter out cycles with non-positive durations or missing values.
    \item \textbf{Model specification:} declare priors, deterministic mean $\mu_i$, and likelihood (\ref{eq:lik}) in PyMC syntax.
    \item \textbf{Sampling:} run NUTS with multiple chains; monitor convergence using $\hat{R}$ and effective sample size.
    \item \textbf{Posterior summarization:} compute posterior distributions of $\tau$ and slope differences $\Delta \beta = \beta_2 - \beta_1$; construct credible intervals.
    \item \textbf{Posterior predictive checks:} draw replicated datasets $z_i^{\text{rep}}$ from $p(z^{\text{rep}} \mid \theta)$ and compare to observed data.
\end{enumerate}

PyMC provides a concise symbolic interface for constructing models while delegating inference to highly optimized numerical backends. Under the hood, PyMC computes gradients of the log-posterior via JAX, enabling efficient exploration of the posterior landscape even when changepoints create mild non-smoothness. The NUTS sampler used here is a variant of Hamiltonian Monte Carlo that eliminates the need to manually specify trajectory lengths; instead, it adaptively determines when to stop the Hamiltonian integration \cite{hoffman2014nuts}. This adaptivity is especially beneficial for non-conjugate models such as changepoint regression.

Model checking is performed using ArviZ, which provides trace plots, rank histograms, kernel density estimates, and divergence diagnostics \cite{arviz2019}. Posterior predictive checks (Fig.~\ref{fig:PPC}) show that replicated draws reproduce the upward trend and variance structure of the observed HI data. In addition, we examined parameter autocorrelation functions and Monte Carlo standard errors, confirming that effective sample sizes are sufficient for stable posterior summaries.

Although PyMC is not intended for deployment directly on embedded systems, it plays a critical role in offline calibration. Once posterior samples are obtained, the posterior medians or full distributions can be compressed into compact lookup tables, small neural surrogates, or polynomial regressors for real-time inference. This mirrors workflows already used in model-based predictive control and impedance-based diagnostics \cite{Goodwin2019,EIS2024Review}.

\subsection{Sampling Details}

For each cell, we run four MCMC chains with 2000 tuning and 2000 sampling iterations per chain, yielding 8000 post-tuning draws. PyMC automatically adapts step sizes and mass matrices during tuning. Diagnostics computed via ArviZ include:

\begin{itemize}
    \item $\hat{R}$ close to 1.0 for all parameters, indicating convergence across chains.
    \item Effective sample sizes (bulk and tail) in the hundreds or thousands for intercepts and slopes.
    \item Zero or negligible number of divergent transitions.
\end{itemize}

\begin{figure}[t]
\centering
\includegraphics[width=\linewidth]{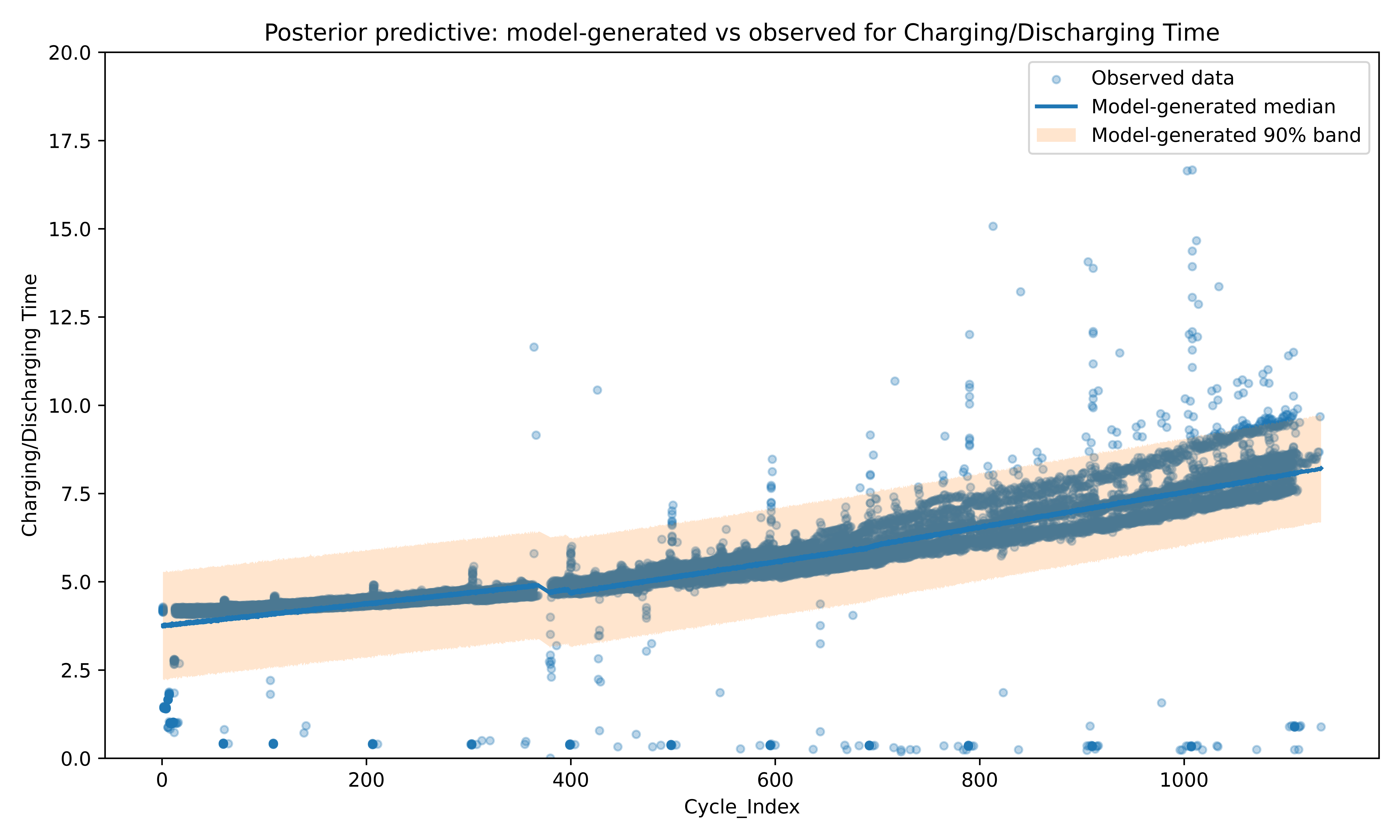}
\caption{Posterior predictive check for a representative cell: observed health indicator (points) and 95\% posterior predictive band (shaded).}
\label{fig:PPC}
\end{figure}

Figure~\ref{fig:PPC} shows a posterior predictive check for one representative cell. The shaded credible band of replicated HI values covers the majority of observed points, indicating that the model’s residual assumptions are compatible with the data.

Although full MCMC would not typically be run on an embedded microcontroller, PyMC-based inference is well suited for offline model fitting, model comparison, and calibration. The resulting posterior summaries or maximum a posteriori (MAP) estimates can then be embedded in a lightweight on-device implementation.

\section{Experimental Results}

We applied the above model to all 14 cell trajectories in the Battery RUL dataset \cite{BatteryRUL2022}. For each cell, we inferred the changepoint location in cycle space and the pre- and post-change slopes of $z_i = \log(\mathrm{HI}_i)$ as functions of $\tilde{t}_i$.

Across cells, posterior distributions of the changepoint $t_{\mathrm{cp}}$ are unimodal and concentrated in the midlife region. Typical posterior medians fall between 380 and 420 cycles, with 95\% highest-density intervals spanning approximately 50–80 cycles. For all cells, the posterior probability that the post-change slope is larger than the pre-change slope, $\Pr(\beta_2 > \beta_1 \mid z)$, exceeds 0.9, supporting the interpretation of accelerated degradation onset.

\begin{figure}[t]
\centering
\includegraphics[width=\linewidth]{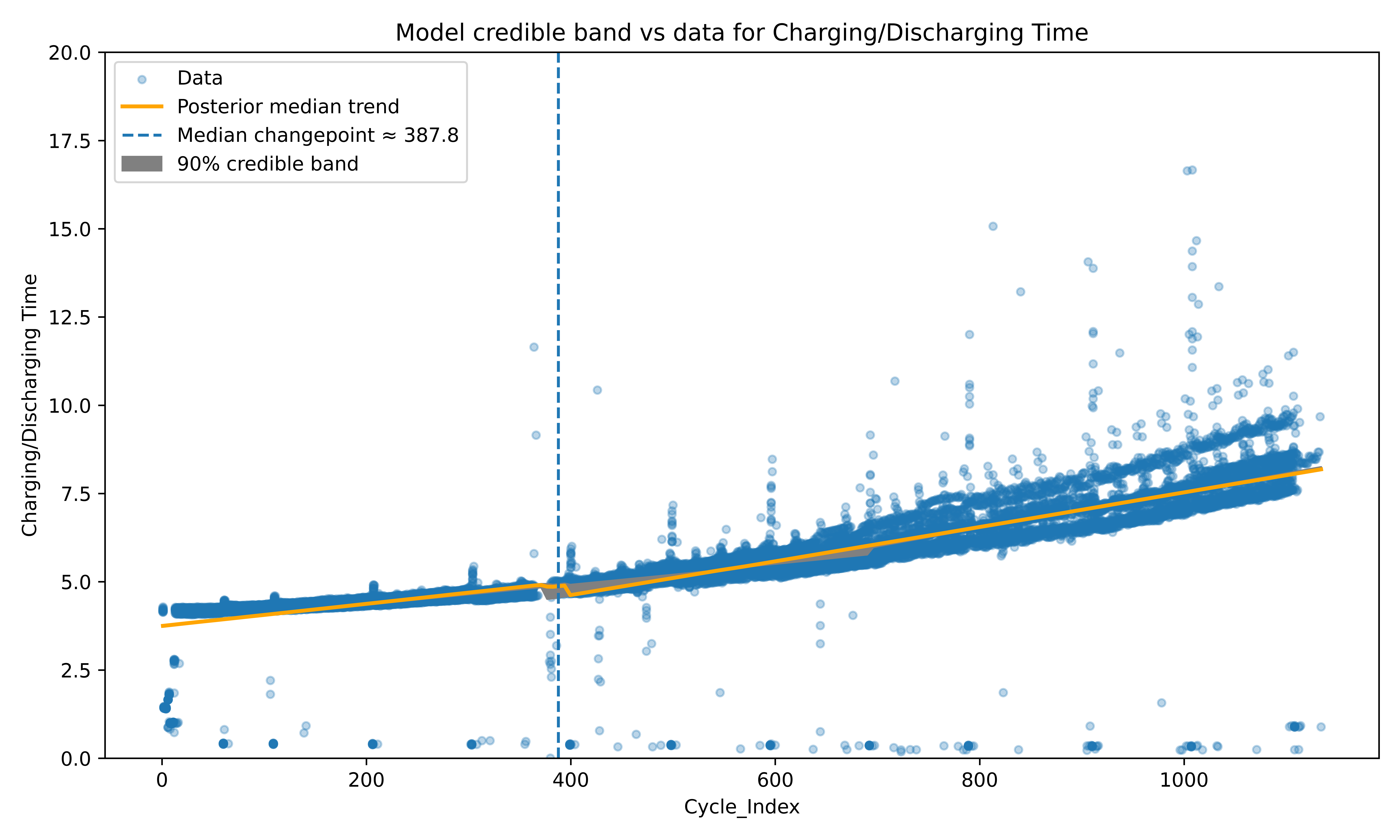}
\caption{Posterior mean trend and 95\% credible band for the log-HI, with the median changepoint indicated by a vertical line.}
\label{fig:Fit}
\end{figure}

Figure~\ref{fig:Fit} illustrates the fitted mean trend and credible band for one representative cell. The model captures an initially flat or slowly varying HI, followed by a steeper increase after the estimated changepoint.

For comparison, simple heuristic methods based on thresholding a smoothed derivative of HI often trigger either too early (sensitive to noise) or too late (over-smoothed), and they provide no uncertainty quantification. In contrast, the Bayesian formulation yields explicit posterior distributions for onset time and change in slope, which are more directly usable in risk-aware BMS logic.

A notable advantage of changepoint-based monitoring is that it directly connects to operational maintenance policies. For example, a BMS may initiate derating when the posterior lower bound of $\tau$ falls below a predefined safety threshold, ensuring high-confidence early warnings. Furthermore, tracking the posterior of the slope difference $\Delta\beta$ enables distinguishing between benign drift and the onset of physically meaningful accelerated degradation.

Future work includes extending the model to multi-output indicators, incorporating physics-informed priors based on electrochemical constraints \cite{Visser2016}, and validating performance under different cycling rates and temperature regimes. Incorporating impedance-derived features \cite{EIS2024Review} or fiber-optic sensing measurements \cite{Fiber2024Review} could also enrich the probabilistic representation, although such enhancements must remain compatible with embedded constraints to preserve the lightweight nature of the approach.

\section{Discussion and Smart-Sensing Implications}

The proposed framework has several properties that make it suitable for smart-sensing applications:

\begin{itemize}
    \item \textbf{BMS compatibility:} the health indicator relies only on cycle-level durations readily available in existing BMS logs; no additional sensors or high-rate storage are required.
    \item \textbf{Interpretability:} the model parameters have clear meanings: pre- and post-change slopes, changepoint location, and residual variability. This aids debugging and regulatory acceptance.
    \item \textbf{Uncertainty quantification:} posterior credible intervals on onset time and slope change enable conservative safety thresholds (e.g., triggering derating when the lower bound of the onset interval enters a given region).
    \item \textbf{Deployability:} while full MCMC is run offline, embedded implementations can use pre-fitted coefficients, approximate filters, or small neural surrogates trained to reproduce posterior summaries.
\end{itemize}

The approach can be extended along several directions: hierarchical changepoint models that share information across cells in a pack; multiple changepoints to capture multi-phase aging; robust likelihoods to handle outliers; and fusion of HI with impedance or temperature-based features \cite{EIS2024Review,Fiber2024Review}. All these can be implemented within the same PyMC-based workflow.

\section{Conclusion}

We have presented a Bayesian single-changepoint framework for detecting the onset of accelerated degradation in lithium-ion batteries using a simple, BMS-friendly health indicator. The model is implemented in PyMC with Hamiltonian Monte Carlo, provides calibrated posterior uncertainty, and shows consistent midlife changepoints on an open 18650-cell dataset. The combination of interpretability, uncertainty quantification, and modest data requirements makes the method attractive for smart-sensing and embedded BMS deployment.

\bibliographystyle{IEEEtran}
\bibliography{refs}

\end{document}